\documentclass[journal,a4paper]{IEEEtran}

\usepackage{enumerate}
\usepackage{multirow}

\usepackage{cite}

\ifCLASSINFOpdf
\else
  
   \usepackage[dvips]{graphicx}
   \DeclareGraphicsExtensions{.eps}
\fi
\usepackage[cmex10]{amsmath}
\usepackage{amsfonts}
\usepackage{amssymb}
\usepackage{supertabular}

\usepackage{scrextend}

\usepackage[tight,footnotesize]{subfigure}

\usepackage{stfloats}
\usepackage[colorlinks=true, citecolor=blue,urlcolor=blue, linkbordercolor={0 0 1}]{hyperref}

\usepackage[none]{hyphenat}

\begin{document}
%
\title{SHAH: Hash Function based on Irregularly Decimated Chaotic Map}
%
%

\author{Mihaela~Todorova, Borislav~Stoyanov, Krzysztof~Szczypiorski, and Krasimir~Kordov
\thanks{Mihaela~Todorova, Borislav Stoyanov, and Krasimir Kordov are with the Department of Computer Informatics, Konstantin Preslavsky University of Shumen, 9712 Shumen, Bulgaria (e-mails: mihaela.todorova@shu.bg, borislav.stoyanov@shu.bg, krasimir.kordov@shu.bg)}
\thanks{Krzysztof~Szczypiorski is with Warsaw University of Technology, Warsaw,  Poland;  Cryptomage  SA,  Wroclaw,  Poland  (e-mail: ksz@tele.pw.edu.pl)}
}

%
%

%
\markboth{}{}
%
%



%
\pagestyle{empty}%
\maketitle%
\thispagestyle{empty}%

\begin{abstract}
In this paper, we propose a novel hash function based on irregularly decimated chaotic map. The hash function called SHAH is based on two Tinkerbell maps filtered with irregular decimation rule.
Exact study has been provided on the novel scheme using distribution analysis, sensitivity analysis, static analysis of diffusion and confusion, and collision analysis. The experimental data show that SHAH satisfied admirable level of security.
\end{abstract}


%
\IEEEpeerreviewmaketitle

\section{Introduction}

%
%
%
%
\IEEEPARstart{D}{uring} recent decades, with the dynamic development of computer science and information technologies, network security tools are becoming increasingly important. 

Decimation sequences play a big part in the area of basic cryptographic primitives.
The output bits are produced by applying a threshold function into a sequence of numbers.
The resulting decimation sequence has good randomness properties.
In \cite{CoppersmithKrawczykMansour1994}, two linear feedback shift registers (LFSRs) and threshold function, named shrinking generator, are used to create a third source of pseudo-random bits. 
A design of a pseudo-random generator based on a single LFSR is proposed in \cite{MeierStaffelbach1994}.
In \cite{KansoSmaoui2009}, a class of irregularly decimated keystream generators, based on 1-D piecewise chaotic map is presented. 
Pseudo-random sequences constructed from the solutions of two Chebyshev maps, filtered by threshold function are presented in \cite{StoyanovCheb}. 
In \cite{HristovMSE2018}, \cite{KordovCircle2015}, \cite{KordovBonchev2017}, \cite{LambicNikolic2017},  \cite{StoyanovLorenzBent}, \cite{StoyanovChebTink}, \cite{StoyanovCircle}, and \cite{StoyanovSzczypiorskiKordovYet2017} new pseudo-random bit generators and software applications
based on discrete chaotic maps, are designed.

A cryptography hash function is a one-way function used for the compression of a plain text of arbitrary length into a secret binary string of fixed-size. The hash function provides the necessary security in authentication and digital signature.

Novel chaos-based hash algorithm, which uses $m$-dimensional Cat map, is proposed in \cite{KwokTang2005}. It is improved in \cite{DengLiXiao2010}, to enhance the influence between the final hash value and the message or key. 

Based on a spatiotemporal chaotic system, a hash construction which has  high performance is designed \cite{RenWangXieYang2009}.
A chaotic look-up table based on a Tent map is used to design a novel 128-bit hash function in \cite{LiXiaoDeng2011}.

A keyed hash algorithm based on a single chaotic skew tent map is constructed in \cite{KansoYahyaouiAlmulla2012}. 
In \cite{KansoGhebleh2013} a new chaotic keyed hash function based on a single 4-dimensional chaotic cat map whose irregular outputs are used to compute a hash value, is designed.

The novel scheme returns a hash value of a fixed length of one of the numbers 128, 160, 256, 512, and 1024. 
In \cite{KansoGhebleh2015}, a 2D generalized Cat map is used to introduce randomness to the computation of the hash value of the input message.

In \cite{YantaoXiang2016}, a circular-shift-based chaotic hash function with variable parameters, is designed. 

A hash function by low 8-bit of 8D hyperchaotic system iterative outputs is proposed in \cite{LinYuLu2017}. 
In \cite{AhmadKhuranaSinghAlSharari2017}, an algorithm for generating secure hash values using a number of chaotic maps is designed .

The aim of the paper is to construct a new hash function based on irregularly decimated chaotic map.

In Section~\ref{sec:TBShrinking-DesignPRBG} we propose a novel pseudo-random bit generator based on two Tinkerbell maps filtered with shrinking rule.
In Section~\ref{sec:TBShrinking-DesignHash} we present the novel hash function SHAH and detailed security analysis is given.
Finally, the last section concludes the paper.

\section{Pseudo-random bit generator based on Irregularly decimated chaotic map} \label{sec:TBShrinking-DesignPRBG}

The work presented in this section was motivated by recent developments in chaos-based pseudo-random generation  \cite{FrancoisGrosgesBarchiesiErra2013}, \cite{StoyanovKordovChirJabri}, and \cite{StoyanovKordovTinkerbell2015}  and with respect of \cite{CoppersmithKrawczykMansour1994}.

\subsection{Proposed Pseudo-random Bit Generation Algorithm}
\begin{figure*}[ht]
	\begin{center}
		\subfigure[]{
			\includegraphics[height=.22\textheight]{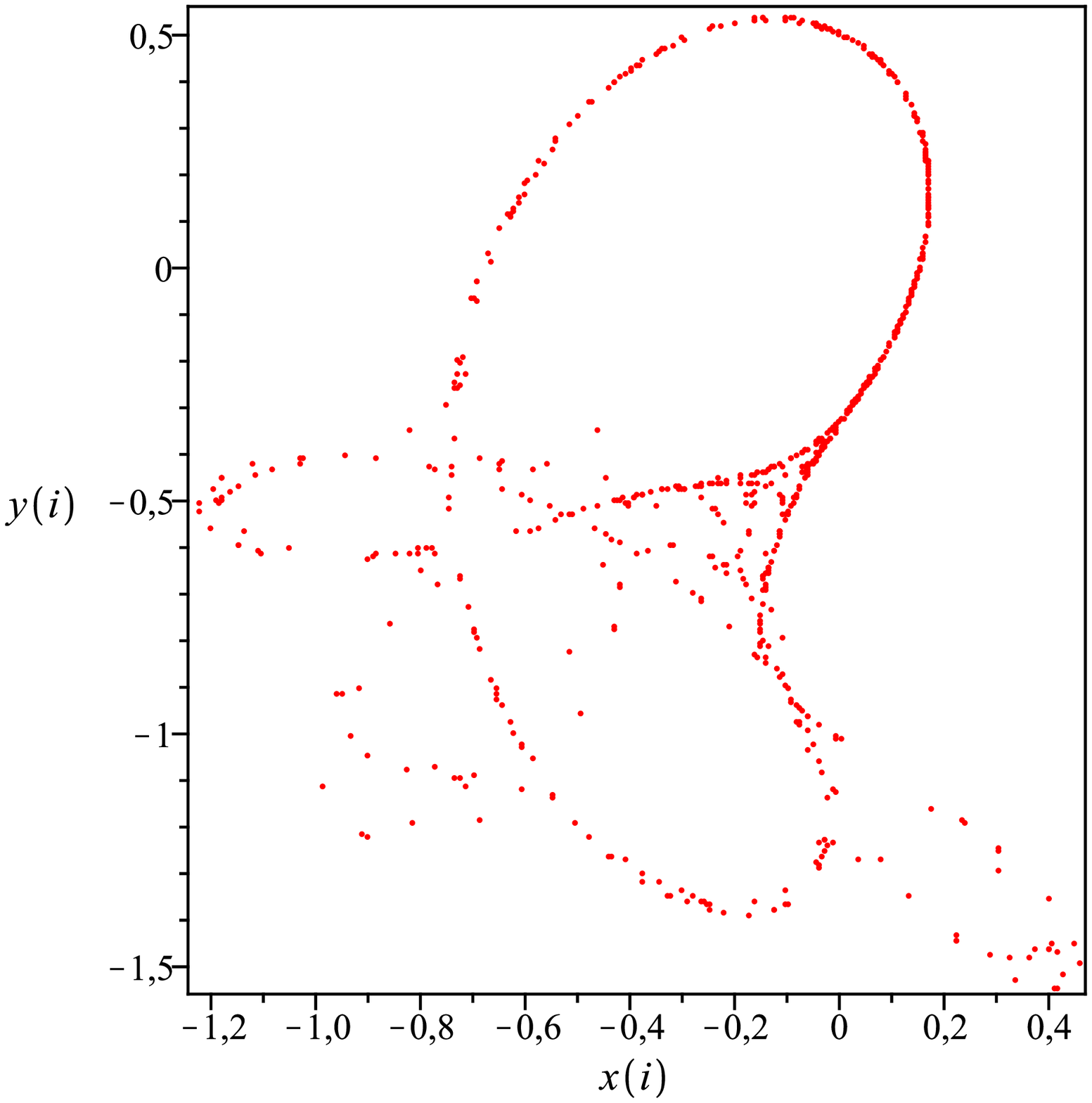}
			\label{fig:TBParamsA}	
		}
		\subfigure[]{
			\includegraphics[height=.22\textheight]{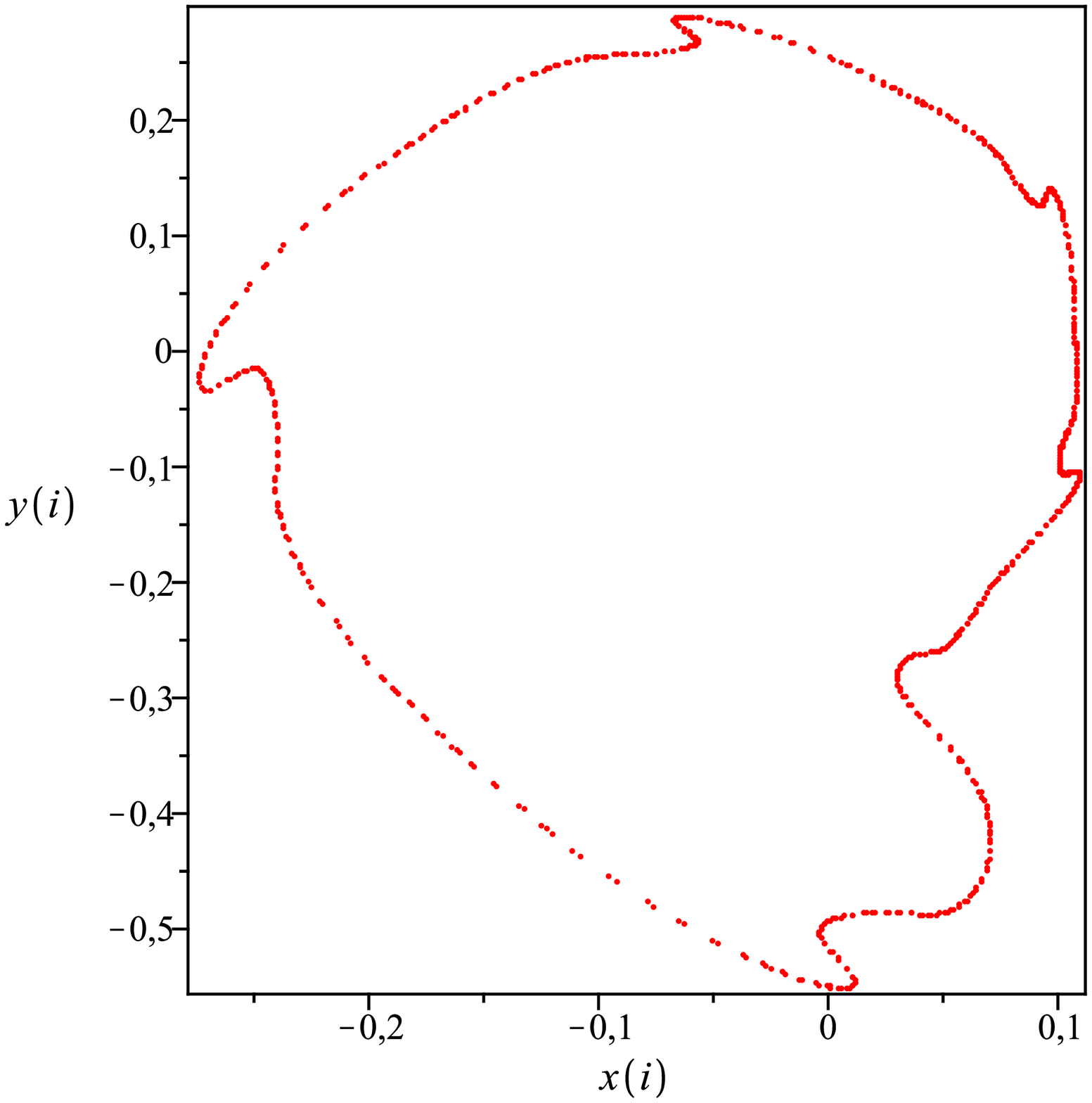}
			\label{fig:TBParamsB}	
		}
		\subfigure[]{
			\includegraphics[height=.22\textheight]{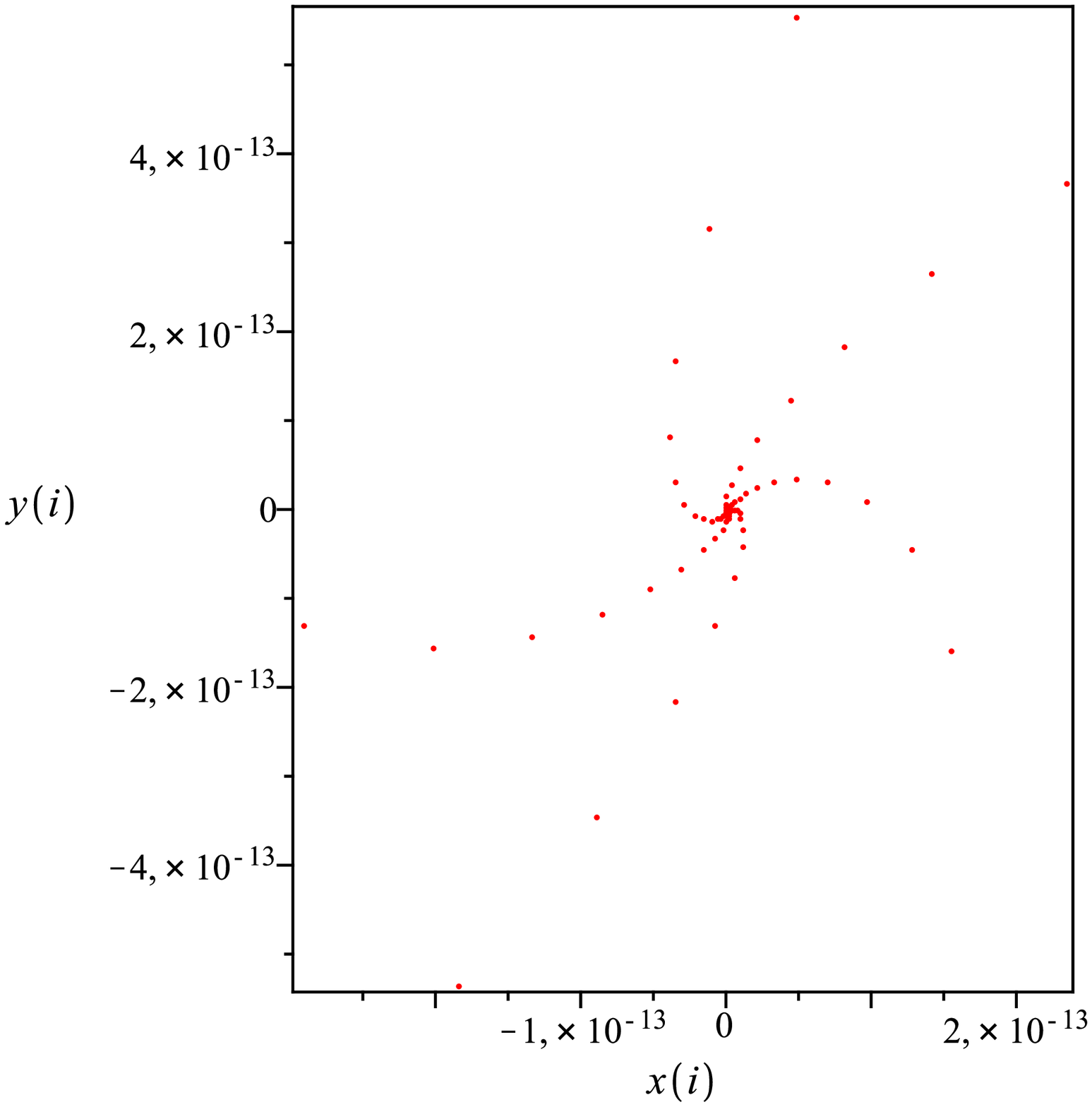}
			\label{fig:TBParamsC}	
		}\\
		\subfigure[]{
			\includegraphics[height=.22\textheight]{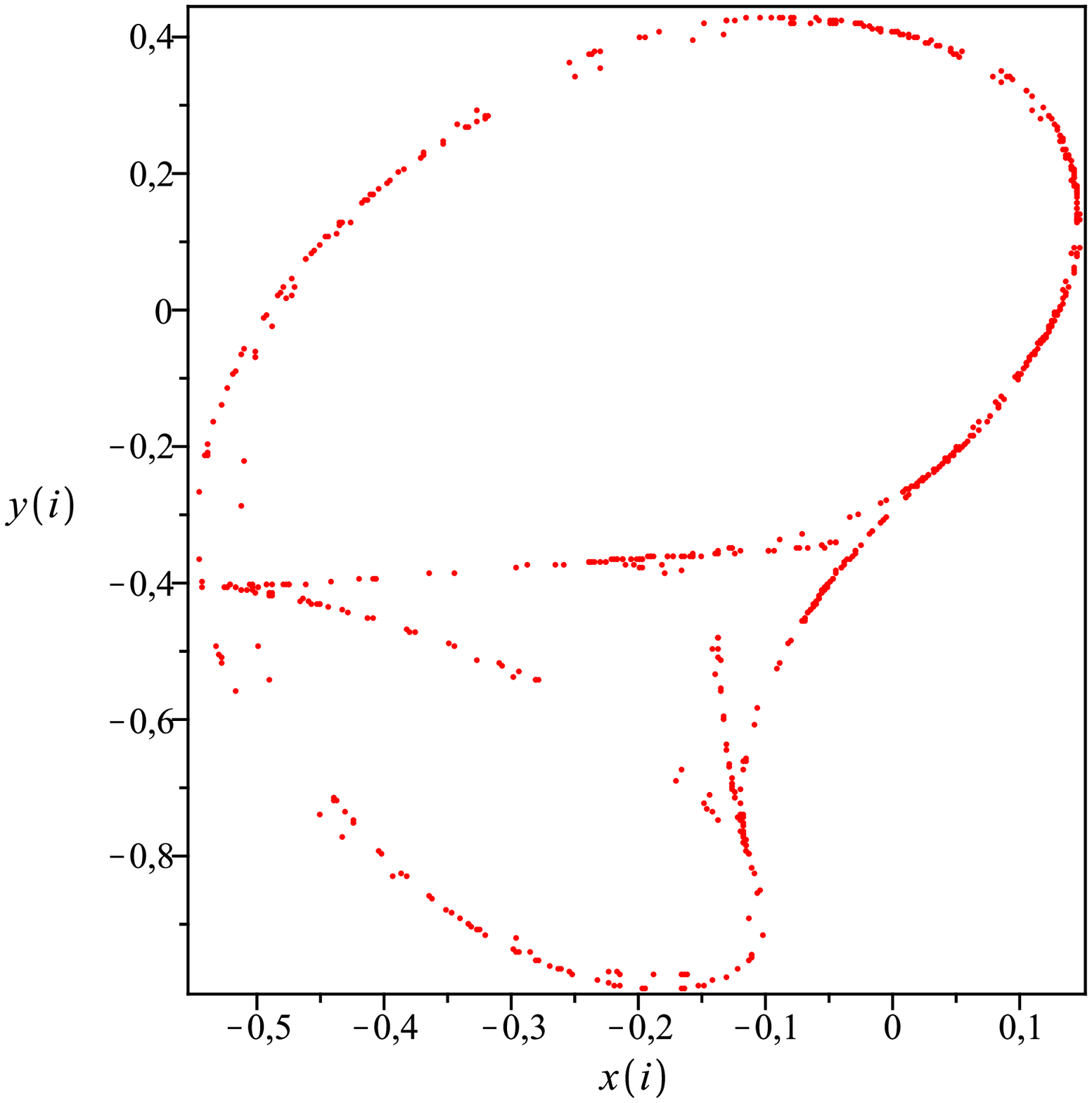}
			\label{fig:TBParamsD}	
		}
		\subfigure[]{
			\includegraphics[height=.22\textheight]{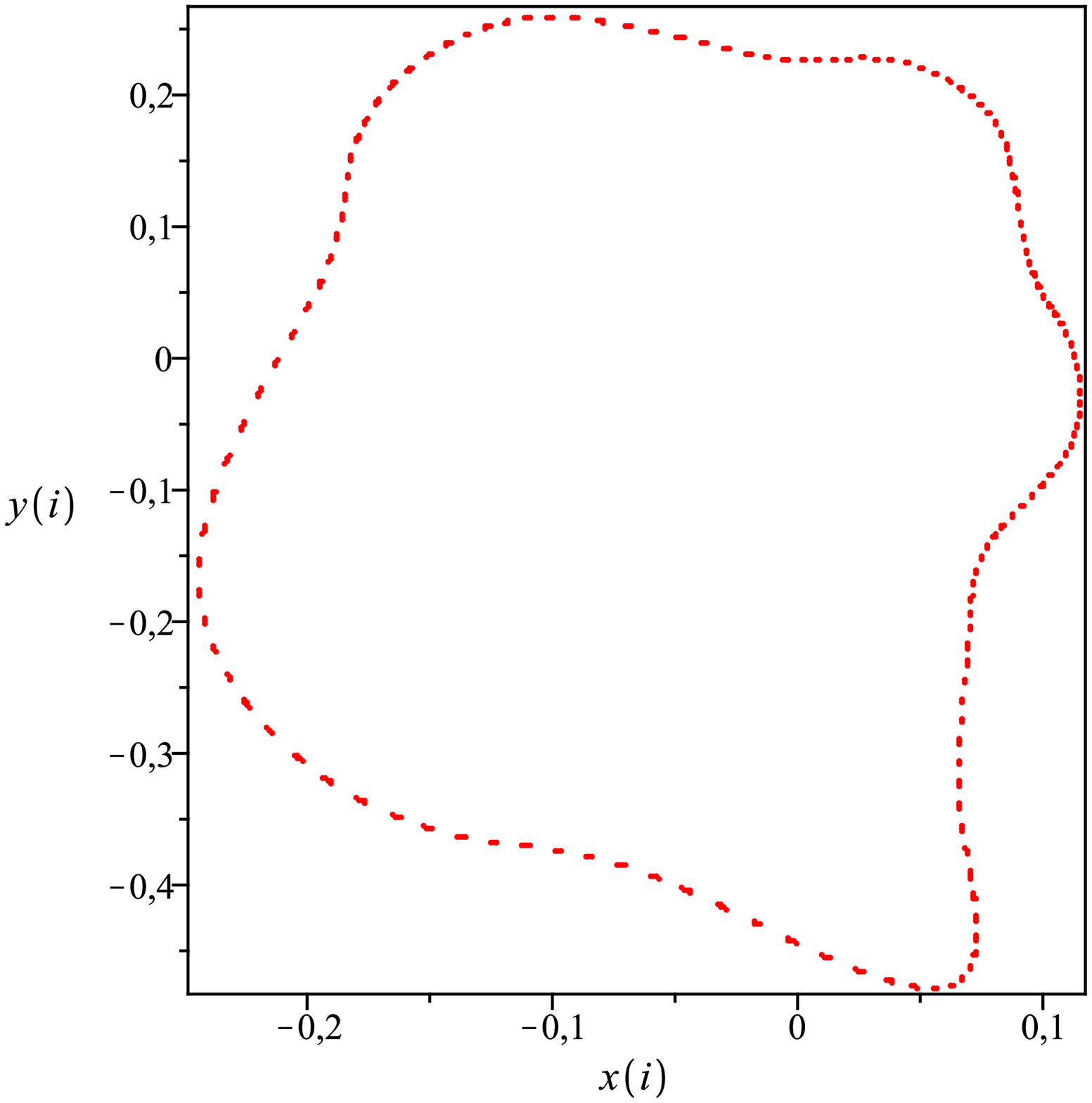}
			\label{fig:TBParamsE}	
		}
		\subfigure[]{
			\includegraphics[height=.22\textheight]{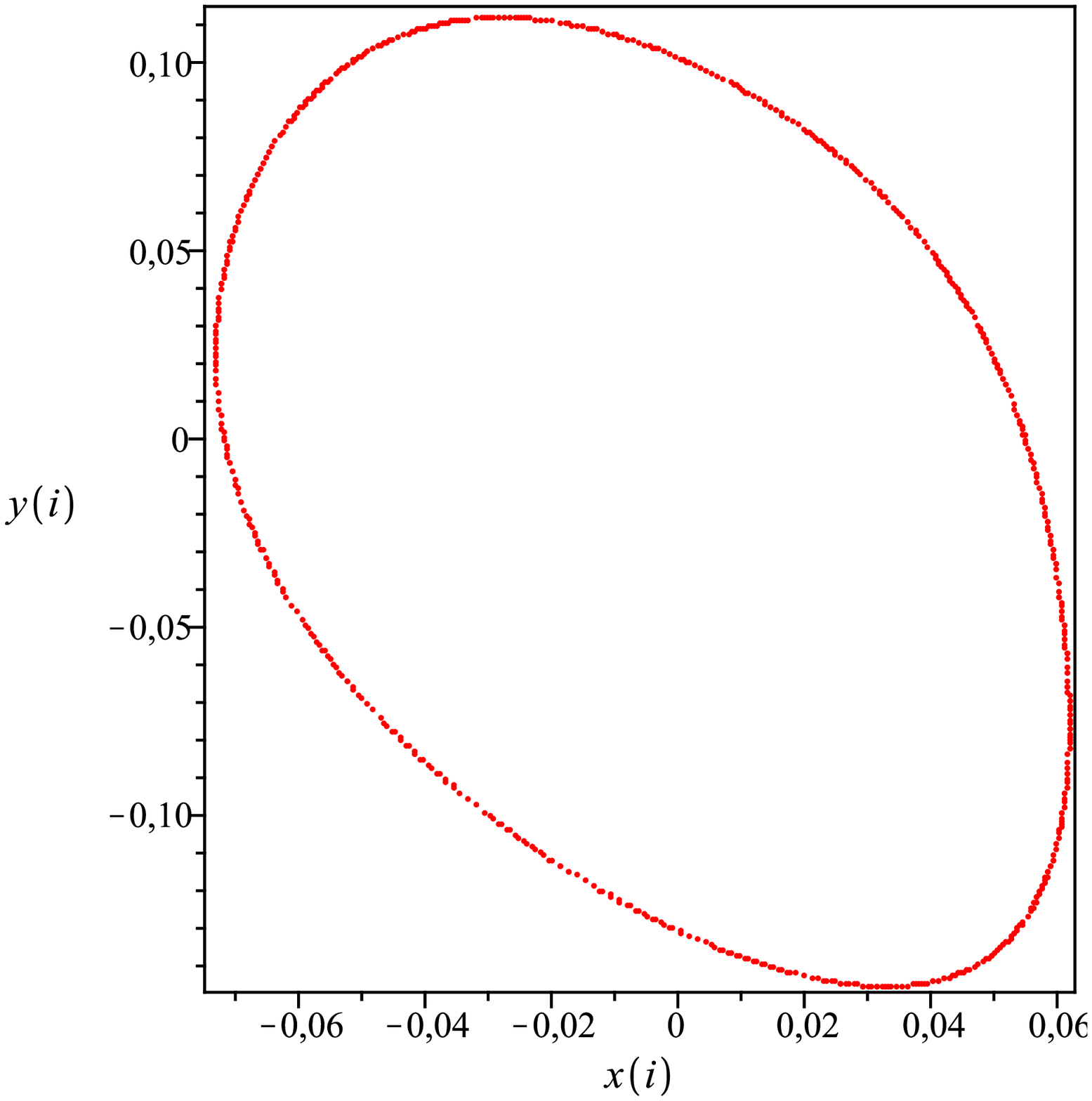}
			\label{fig:TBParamsF}	
		}
	\end{center}	
	\caption{The Tinkerbell map of Eq.(\ref{TBMapEq}). This is a plot of $(x_{i},y_{i})$, for $i=0..100000$. 
({\bf a}) Tinkerbell map with $c_{1}=0.9$, $c_{2}=-0.6013$, $c_{3}=2.0$, and $c_{4}=0.50$, 
({\bf b}) Tinkerbell map with $c_{1}=0.3$, $c_{2}=-0.6013$, $c_{3}=2.0$, and $c_{4}=0.50$, 
({\bf c}) Tinkerbell map with $c_{1}=0.9$, $c_{2}=-0.6013$, $c_{3}=2.0$, and $c_{4}=-0.4$, 
({\bf d}) Tinkerbell map with $c_{1}=0.9$, $c_{2}=-0.6013$, $c_{3}=2.0$, and $c_{4}=0.4$, 
({\bf e}) Tinkerbell map with $c_{1}=0.3$, $c_{2}=-0.6013$, $c_{3}=2.0$, and $c_{4}=0.4$, and 
({\bf f}) Tinkerbell map with $c_{1}=-0.3$, $c_{2}=-0.6013$, $c_{3}=2.0$, and $c_{4}=0.5$.}
	\label{fig:TBParams} 
\end{figure*}

The Tinkerbell map \cite{AlligoodSauerYorke1996} is a discrete-time dynamical system given by:
\begin{equation}\label{TBMapEq}
	\begin{aligned}
		&x_{n+1}=x_{n}^{2}-y_{n}^{2}+c_{1}x_{n}+c_{2}y_{n} \\
		&y_{n+1}=2x_{n}y_{n}+c_{3}x_{n}+c_{4}y_{n} \ .
	\end{aligned}
\end{equation}
The map depends on the four paremeters $c_{1}$, $c_{2}$, $c_{3}$, and $c_{4}$. The Tinkerbell map with different values of the parameters is illustrated in Figure \ref{fig:TBParams}.

The shrinking generator \cite{CoppersmithKrawczykMansour1994} uses two sequences of pseudo-random bits (\textbf{a} and \textbf{s}) to create a third source (\textbf{z}) of pseudo-random bits which includes those bits $a_{i}$ for which the corresponding $s_{i}$ is 1. Other bits from the first sequence are decimated.
 
We propose a novel pseudo-random number scheme which irregularly decimates the solutions of two Tinkerbell maps by using the shrinking rule \cite{CoppersmithKrawczykMansour1994}. We used the following parameters $a=0.9$, $b=-0.6013$, $c=2.0$ and $d=0.50$. 
The novel generator is based on the following equations:
\begin{equation}\label{TBMapSo}
	\begin{aligned}
		&x_{1,n+1}=x_{1,n}^{2}-y_{1,n}^{2}+c_{1}x_{1,n}+c_{2}y_{1,n} \\
		&y_{1,n+1}=2x_{1,n}y_{1,n}+c_{3}x_{1,n}+c_{4}y_{1,n} \\ 
		&x_{2,m+1}=x_{2,m}^{2}-y_{2,m}^{2}+c_{1}x_{2,m}+c_{2}y_{2,m} \\
		&y_{2,m+1}=2x_{2,m}y_{2,m}+c_{3}x_{2,m}+c_{4}y_{2,m} \ ,	
	\end{aligned}
\end{equation}
where initial values $x_{1,0}, y_{1,0}, x_{2,0}$ and $y_{2,0}$ are used as a key.
\begin{enumerate} [Step 1:]
	{\item The initial values $x_{1,0}, y_{1,0}, x_{2,0}$ and $y_{2,0}$ of the two Tinkerbell maps from Eqs. (\ref{TBMapSo}) are determined.}
	{\item The first and the second Tinkerbell maps from Eqs. (\ref{TBMapSo}) are iterated for $M$ and $N$ times, respectively, to avoid the harmful effects of transitional procedures, where $M$ and $N$ are different constants.}
	{\item The iteration of the Eqs. (\ref{TBMapSo}) continues, and as a result, two real fractions $y_{1,n}$ and $y_{2,m}$ are generated and preprocessed as follows:
		\begin{equation}\label{Preproc}
			\begin{aligned}
				&a_{i}= abs(mod(integer(y_{1,n}\times 10^{9}),2) \\		
				&s_{i}= abs(mod(integer(y_{2,m}\times 10^{9}),2),
			\end{aligned}
		\end{equation}  
		
		where $abs(x)$ returns the absolute value of $x$, $integer(x)$ returns the the integer part of $x$, truncating the value at the decimal point, $mod(x,y)$ returns the reminder after division.}
		{\item Apply the shrinking rule \cite{CoppersmithKrawczykMansour1994} to the values $(a_{i},z_{i})$ and produce the output bit.}
		{\item Return to Step 3 until the bit stream limit is reached.}
\end{enumerate}

The novel generator is implemented by software simulation in C++ language, using the values: 
$x_{1,0}=$ $-0.423555643379287$, $y_{1,0}=$ $-0.762576287931311$, $M=N=3500$, 
$x_{2,0}=$ $-0.276976682878721$, and $y_{2,0}=$ $-0.348339839900213$.

\subsection{Key space calculation}
The key space is the set of all input values that can be used as a seed of the pseudo-random bit generation steps. The proposed generator has four input parameters $x_{1,0}$, $y_{1,0}$, $x_{2,0}$, and $y_{2,0}$. According to \cite{IEEE}, the computational precision of the
64-bit double-precision number is about $10^{-15}$, thus the key space is more than $2^{199}$. The proposed pseudo-random generator is secure against exhaustive key search\cite{AlvarezLi}.
Moreover, the initial iteration numbers $M$ and $N$ can also be used as a part of the key space.

\subsection{Statistical tests}
In order to measure randomness of the sequences of bits produced by the new pseudo-random number algorithm, 
we used the statistical applications NIST \cite{BS:NIST}, DIEHARD \cite{DIE}, and ENT \cite{ENT}.

The NIST statistical test package (version 2.1.1) includes 15 tests, which focus on the randomness of binary sequences produced by either hardware or software-based bit generators. These tests are: frequency (monobit), block-frequency, cumulative sums, runs, longest run of ones, rank, Fast Fourier Transform (spectral), non-overlapping templates, overlapping templates, Maurer's "Universal Statistical", approximate entropy, random excursions, random-excursion variant, serial, and linear complexity.

For the NIST tests, we generated $10^{3}$ different binary sequences of length $10^{6}$ bits. The results from the tests are given in Table \ref{tab:NIST}.
\begin{table}
\caption{NIST test suite results.}
\label{tab:NIST}
\footnotesize
\centering
\begin{tabular}{lrr}
\hline \hline
	\tablehead{1}{\textbf{NIST}}
  & \multicolumn{2}{c}{\textbf{SHAH Algorithm}}\\ 
\cline{2-3}
	\tablehead{1}{\textbf{statistical test}}
  & \tablehead{1}{\textit{\textbf{P-value}}}
  & \tablehead{1}{\textbf{Pass rate}}\\ 
\hline
Frequency (monobit) 		& 0.869278  & 981/1000 \\
Block-frequency			& 0.548314  & 985/1000 \\
Cumulative sums (Reverse)	& 0.790621  & 983/1000 \\
Runs						& 0.610070  & 990/1000 \\
Longest run of Ones			& 0.439122  & 984/1000 \\
Rank						& 0.467322  & 989/1000 \\
FFT						& 0.058612  & 988/1000 \\
Non-overlapping templates	& 0.519879  & 991/1000 \\
Overlapping templates		& 0.510153  & 982/1000 \\
Universal					& 0.159910  & 989/1000 \\
Approximate entropy			& 0.616305  & 991/1000 \\
Random-excursions			& 0.641892  & 588/594 \\
Random-excursions Variant	& 0.495265  & 589/594 \\
Serial 1					& 0.614226  & 989/1000 \\
Serial 2					& 0.151190  & 985/1000 \\
Linear complexity			& 0.620465  & 990/1000 \\
\hline \hline
\end{tabular}
\end{table}
The minimum pass rate for each statistical test with the exception of the random excursion (variant) test is approximately = 980 for a sample size = $10^{3}$ binary sequences.
The minimum pass rate for the random excursion (variant) test is approximately = 580 for a sample size = 594 binary sequences.
The proposed pseudo-random bit generator passed successfully all the NIST tests.

The DIEHARD application \cite{DIE} is a set of 19 statistical tests: birthday spacings, overlapping 5-permutations, binary rank (31 x 31), binary rank (32 x 32), binary rank (6 x 8), bitstream, overlapping-pairs-sparse-occupancy, overlapping-quadruples-sparse-occupancy, DNA, stream count-the-ones, byte-count-the-ones, parking lot, minimum distance, 3D spheres, squeeze, overlapping sums, runs (up and down), and craps. The tests return $P-values$, which should be uniform in [0,1), if the input stream contains pseudo-random numbers. The $P-values$ are obtained by $p=F(y)$, where $F$ is the assumed distribution of the sample random variable $y$, often the normal distribution.
The novel pseudo-random bit algorithm passed successfully all DIEHARD tests, Table \ref{tab:DIEHARD}.
\begin{table}
\caption{DIEHARD statistical test results.}
\label{tab:DIEHARD}
\footnotesize
\centering
\begin{tabular}{lc}
\hline \hline
	\tablehead{1}{\textbf{DIEHARD}} 
  & \tablehead{1}{\textbf{SHAH Algorithm}}\\ 
	\tablehead{1}{\textbf{statistical test}}
  & \tablehead{1}{\textit{\textbf{P-value}}}\\ 
\hline
Birthday spacings         & 0.513830  \\
Overlapping 5-permutation & 0.927974  \\
Binary rank (31 x 31)     & 0.890892  \\
Binary rank (32 x 32)     & 0.609788  \\
Binary rank (6 x 8)       & 0.486987  \\
Bitstream                 & 0.662411  \\
OPSO                      & 0.618526  \\
OQSO                      & 0.445982  \\
DNA                       & 0.526710  \\
Stream count-the-ones     & 0.299022  \\
Byte count-the-ones       & 0.546796  \\
Parking lot               & 0.574512  \\
Minimum distance          & 0.115118  \\
3D spheres                & 0.527506  \\
Squeeze                   & 0.678411  \\
Overlapping sums          & 0.556561  \\
Runs up                   & 0.543542  \\
Runs down                 & 0.438540  \\
Craps                     & 0.272223  \\
\hline \hline
\end{tabular}
\end{table}

The ENT software \cite{ENT} includes 6 tests of pseudo-random sequences: entropy, optimum compression, ${\chi}^2$ distribution, arithmetic mean value, Monte Carlo value for $\pi$, and serial correlation coefficient. Sequences of bytes are stored in files. The application outputs the results of those tests. We tested output sequences of 125,000,000 bytes of the novel pseudo-random bit generation scheme.
The novel pseudo-random bit generation algorithm passed successfully all ENT test, Table \ref{tab:ENT}.

\begin{table}
\caption{ENT statistical test results.}
\label{tab:ENT}
\centering
\begin{tabular}{ll}
\hline \hline
	\tablehead{1}{\textbf{ENT}}
  & \tablehead{1}{\textbf{SHAH Algorithm}}\\ 
	\tablehead{1}{\textbf{statistical test}}
  & \tablehead{1}{\textbf{results}}\\ 
\hline
Entropy                   & 7.999998 bits per byte       \\
Optimum compression       & OC would reduce the size of  \\
					      & this 125000000 byte file     \\
					      & by 0 $\%$.                   \\
${\chi}^2$ distribution   & For 125000000 samples is     \\
					      & 278.28, and randomly would   \\
					      & exceed this value 15.15 $\%$ \\					      
					      & of the time.                 \\
Arithmetic mean value     & 127.5015  (127.5 = random)   \\
Monte Carlo $\pi$ estim.  & 3.141354290 (error 0.01 $\%$)\\
Serial correl. coeff.     & 0.000115                    \\
					      & (totally uncorrelated = 0.0) \\
\hline
\end{tabular}
\end{table}

Based on the good test results, we can conclude that the novel pseudo-random bit generation algorithm has satisfying statistical properties and provides acceptable level of security.

\section{Hash Function based on Irregularly Decimated Chaotic Map} \label{sec:TBShrinking-DesignHash}

\subsection{Proposed Hash Function based on Irregularly Decimated Chaotic Map}
In this section, we construct a keyed hash function named SHAH based on a irregularly decimated chaotic map.
Let $n$ be the bit length of the final hash value.
The parameter $n$ usually supports five bit lengths, 128, 160, 256, 512, and 1024 bits.
We consider input message $M'$ with arbitrary length. 

The novel hash algorithm SHAH consists of the following steps:
\begin{enumerate} [Step 1:]
	
	{\item Convert the input message $M$ to binary sequence using ASCII table. }

	{\item  The input message $M$ is padded with a bit of one, and then append zero bits to obtain a message $M'$ whose length is $m$, a multiple of $n$.}

	{\item  The novel pseudo-random bit generation algorithm (Section \ref{sec:TBShrinking-DesignPRBG}) based on two Tinkerbell maps filtered with shrinking rule is iterated many times, getting $m$ bits, $m$-sized vector $P$.}

	{\item The $m$-sized vectors ${M'}$ and $P$ are combined in a new $m$-sized vector, $N$, using XOR operation.}

	{\item The vector $N$ is split into $p$ blocks, $N_1, N_2, ...,N_p$, each of length $n$ and $m=np$ is the total length of the vector ${N}$.}	

	{\item  A temporary $n$-sized vector ${T}$ is obtained by
${T}=N_1 \oplus N_2 \oplus \cdots \oplus N_p $.}

	{\item  The bits from the temporary vector ${T}$ are processed one by one sequentially. If the current bit $t_i$ is 1 then update $t_i=t_i \oplus s$, where $s$ is the next bit from the novel pseudo-random generator based on Tinkerbell function (Section \ref{sec:TBShrinking-DesignPRBG}).}

	{\item Another $n$-sized temporary vector ${U}$ is taken and all the elements are initialized to 0s.}
	{\item The bits from the vector ${T}$ are processed again one by one sequentially. 
	If the current bit $t_i$ is 1, the vector ${U}$ is XOR-ed with the next $n$ bits from the novel pseudo-random generator based on Tinkerbell function (Section \ref{sec:TBShrinking-DesignPRBG}).
	If the current bit $t_i$ is 0, the matrix ${U}$ is bitwise rotated left by one bit position.}
	{\item The final hash value  is obtained by $H={T} \oplus {U}$.}
\end{enumerate}

The designed SHAH algorithm is implemented in C++ programming language.

\subsection{Distribution Analysis}
\label{sec:DistributionAnalysis}

\begin{figure*}[ht]
\begin{center}
\subfigure[]{
	\includegraphics[height=.25\textheight]{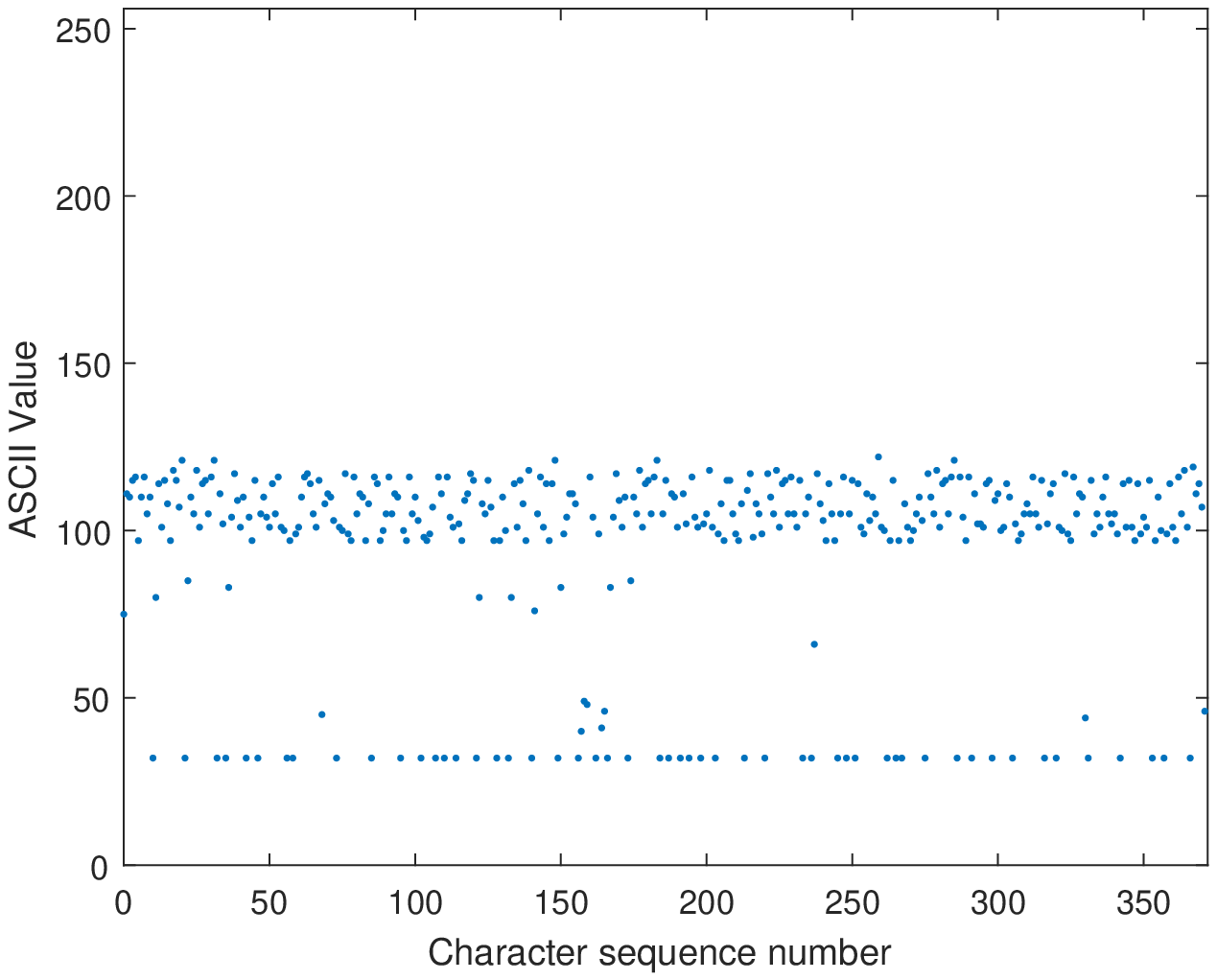}
	\label{fig:shah-distrib-ascii-c1}	
}
\subfigure[]{
	\includegraphics[height=.25\textheight]{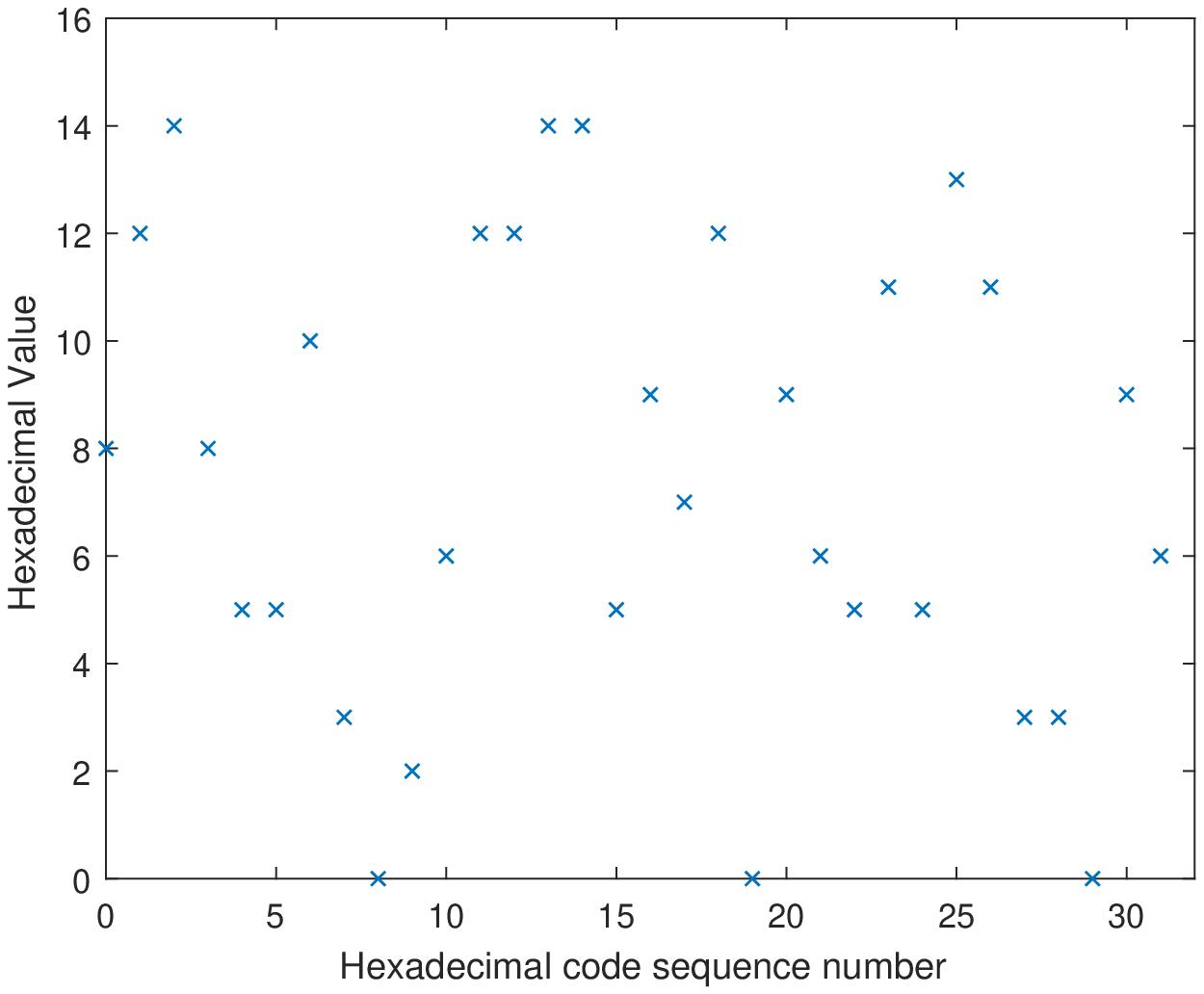}
	\label{fig:shah-distrib-hash-c1}	
}
\\
\subfigure[]{
	\includegraphics[height=.25\textheight]{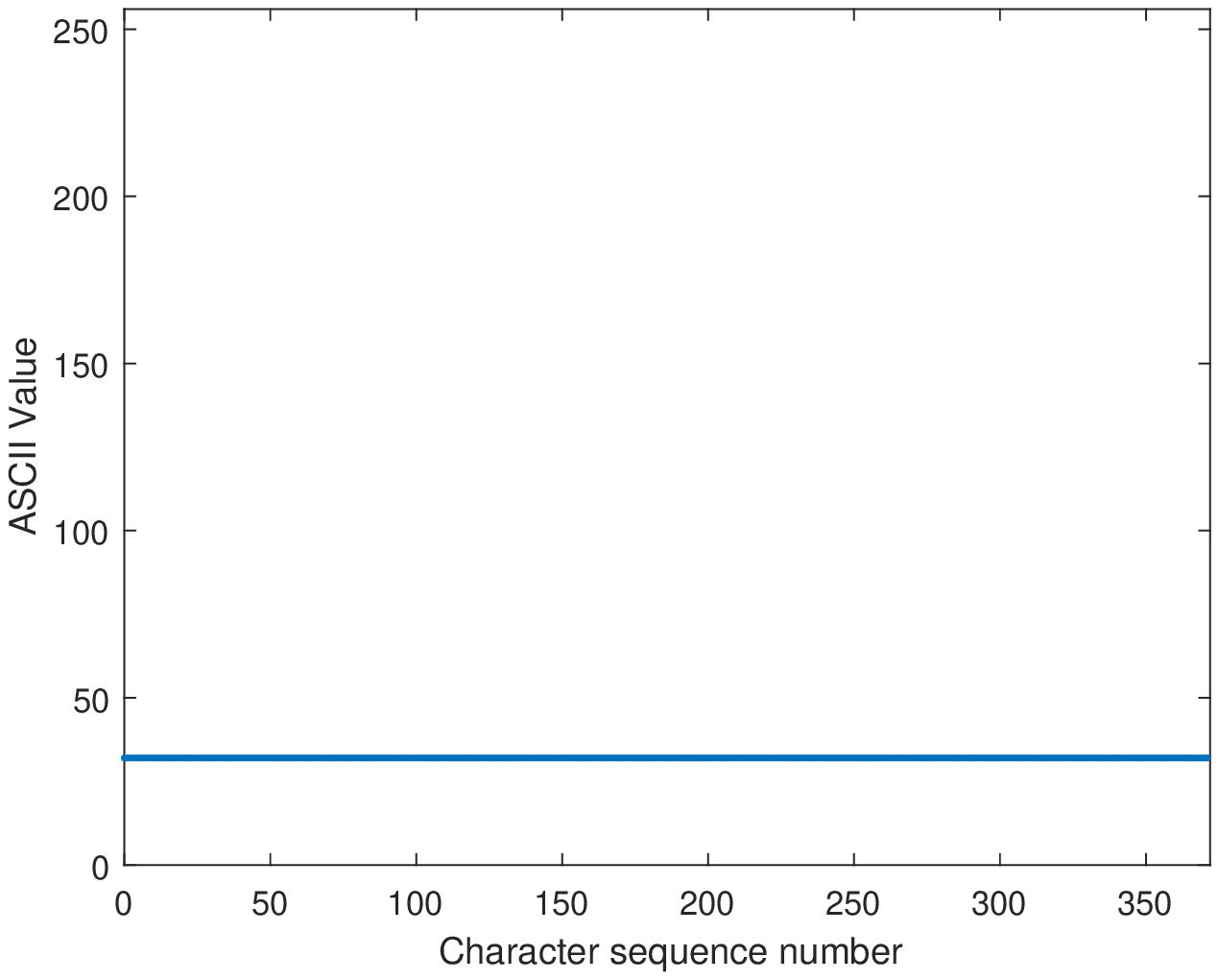}
	\label{fig:shah-distrib-blank-c0}	
}
\subfigure[]{
	\includegraphics[height=.25\textheight]{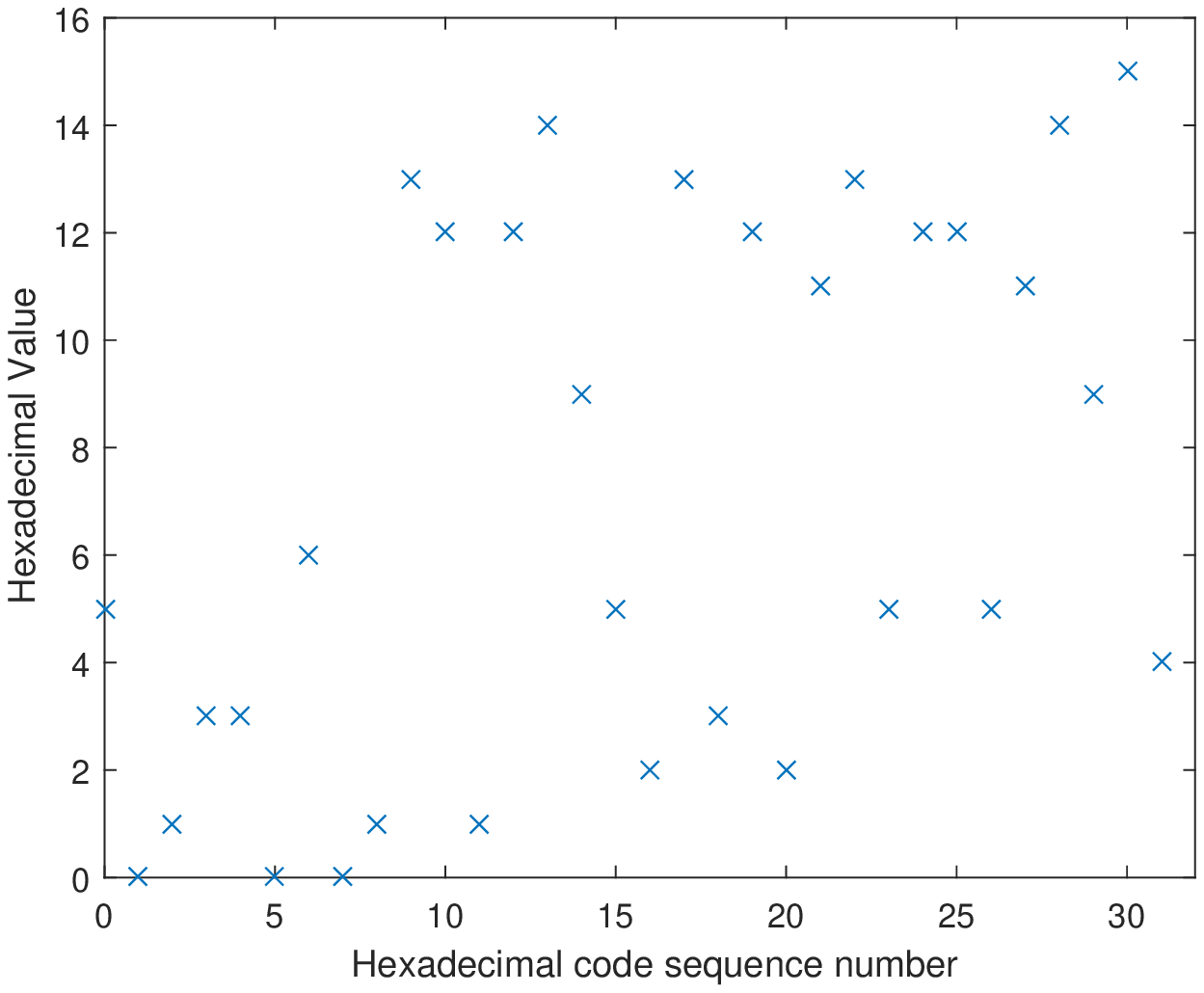}
	\label{fig:shah-distrib-hash-c0}	
}
\end{center}	
	\caption{Distribution of input message and corresponding hash value.}
	\label{fig:shah-distrib}
\end{figure*}

In general, a typical property of a hash value is to be uniformly distributed in the compressed range.
Note that the length of the hash value is set as 128.
Simulation experiments are done on the following paragraph of message:
\textit{\begin{addmargin}[1em]{1em}
{Konstantin Preslavsky University of Shumen has inherited a centuries-long educational tradition dating back to the famous Pliska and Preslav Literary School (10th c). Shumen University is one of the five classical public universities in Bulgaria it is recognized as a leading university that offers modern facilities for education, scientific researches and creative work.}
\end{addmargin}
}
With the chosen input message, the SHAH hash value is calculated.
The ASCII code distribution of input message and the corresponding hexadecimal
hash value are shown in Fig. \ref{fig:shah-distrib-ascii-c1} and \ref{fig:shah-distrib-hash-c1}.
Another input message with the same length but all of blank spaces, is generated.
The ASCII code distribution of the blank-spaced input message and the corresponding hexadecimal hash value are shown in Fig. \ref{fig:shah-distrib-blank-c0} and \ref{fig:shah-distrib-hash-c0}.
The SHAH hash plots, \ref{fig:shah-distrib-hash-c1} and \ref{fig:shah-distrib-hash-c0}, are uniformly distributed in compress range even under exceptionally cases.

\subsection{Sensitivity Analysis}
In order to demonstrate the sensitivity of the proposed keyed hash function to the input message and security key space, hash simulation experiments have been performed under the following 9 cases:

Case 1: The input message is the same as the one in Section \ref{sec:DistributionAnalysis};

Case 2: Change the first character \textit{'K'} in the input message into \textit{'k'}.

Case 3: Change the number \textit{'10'} in the input message to \textit{'11'}.

Case 4: Change the word \textit{'School'} in the input message to \textit{'school'}.

Case 5: Change the comma \textit{','} in the input message to \textit{'.'}.

Case 6: Add a blank space at the end of the input message.

Case 7: Change the word \textit{'recognized'} in the input message to \textit{'recognize'}.

Case 8: Subtracts $1\times 10^{-15}$ from the input key value $x_{1,0}$.

Case 9: Adds $1\times 10^{-15}$ to the input key value $y_{2,0}$.

The respective 128-bit hash values in hexadecimal number system are the following:

Case 1: 8CE855A3026CCEE597C0965B5DB33096

Case 2: 8F0E33DA59B5B1114F9A1570EB466C24

Case 3: 11DEA1F51379EC2B429325D16FD5354C

Case 4: 6B834AC8D36B74EFAD0C6B8AAEA008BF

Case 5: 81704BC6412FF4E24AF09E570AB4D9DE

Case 6: A2B7D2EAC687D2953551AE2621720ADF

Case 7: E1D9A0BAA6184264481A25D08BEFF110

Case 8: 780378A8FE0011DBD81CE035414907F0

Case 9: C27AB518A87B8E3D6C46504814BF7940

\begin{figure}[ht]
\centering
\subfigure[]{
	\includegraphics[height=.039\textheight]{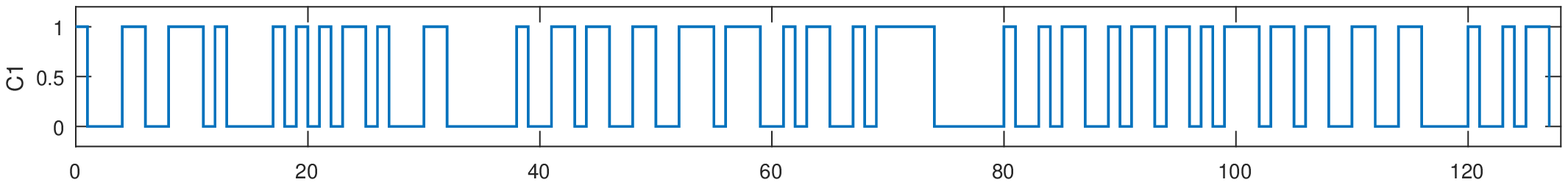}
	\label{fig:shah-sens-c1}	
}
\\
\subfigure[]{
	\includegraphics[height=.039\textheight]{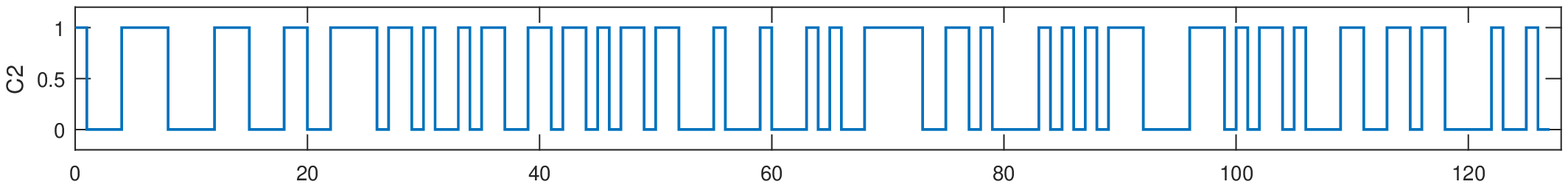}
	\label{fig:shah-sens-c2}	
}
\\
\subfigure[]{
	\includegraphics[height=.039\textheight]{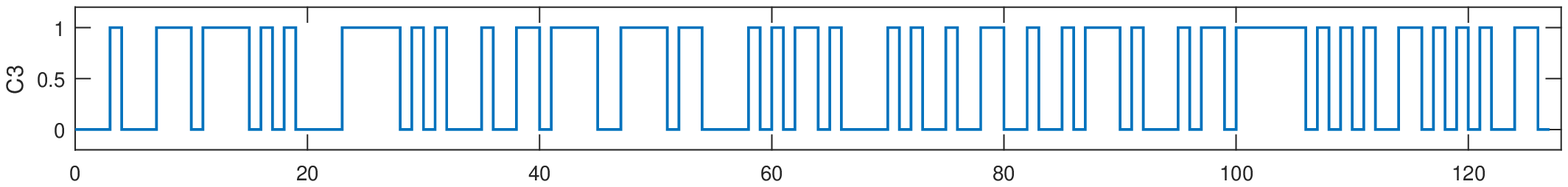}
	\label{fig:shah-sens-c3}	
}
\\
\subfigure[]{
	\includegraphics[height=.039\textheight]{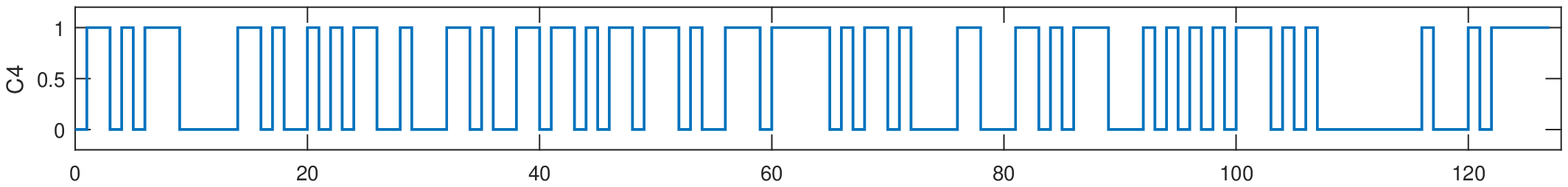}
	\label{fig:shah-sens-c4}	
}
\\
\subfigure[]{
	\includegraphics[height=.039\textheight]{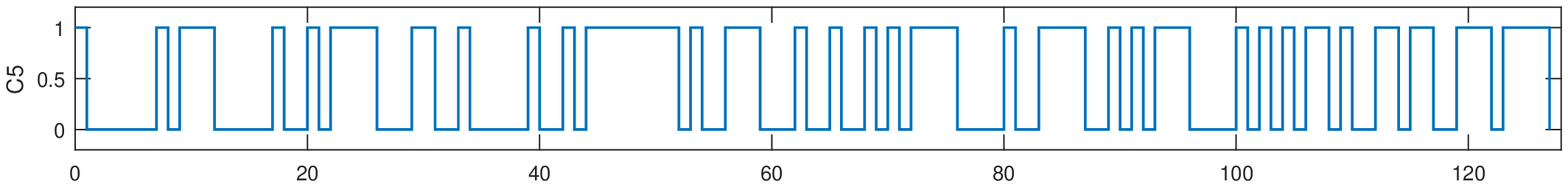}
	\label{fig:shah-sens-c5}	
}
\\
\subfigure[]{
	\includegraphics[height=.039\textheight]{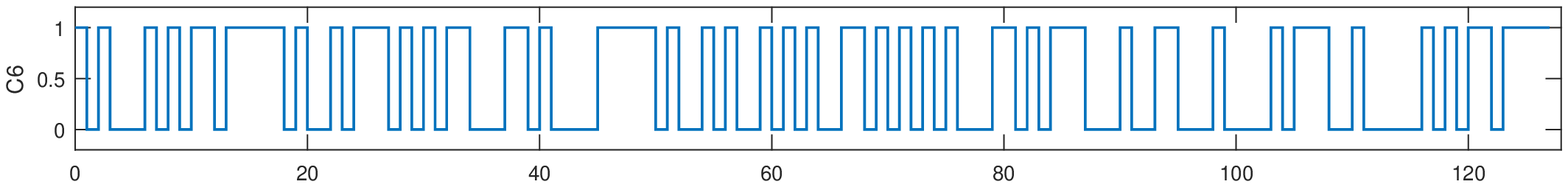}
	\label{fig:shah-sens-c6}	
}
\\
\subfigure[]{
	\includegraphics[height=.039\textheight]{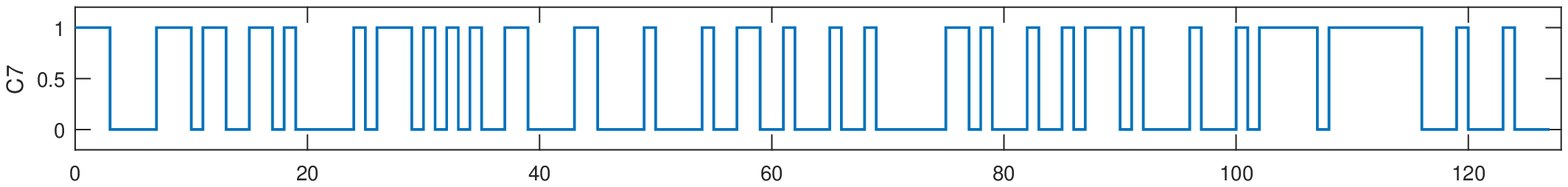}
	\label{fig:shah-sens-c7}	
}
\\
\subfigure[]{
	\includegraphics[height=.039\textheight]{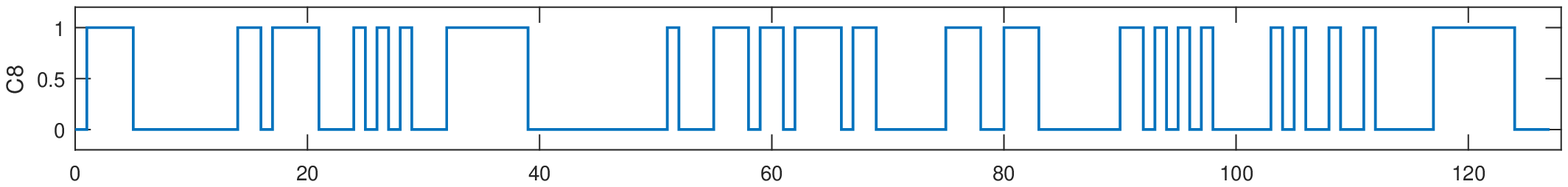}
	\label{fig:shah-sens-c8}	
}
\\
\subfigure[]{
	\includegraphics[height=.039\textheight]{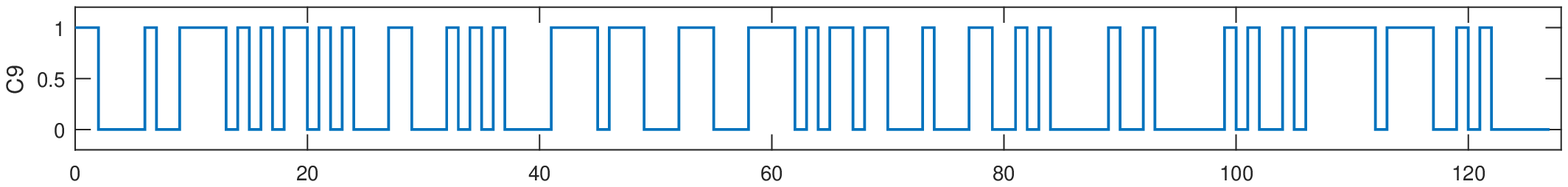}
	\label{fig:shah-sens-c9}	
}

	\caption{128-bit hash values of the input messages under nine different cases: ({\bf a}) Case 1, ({\bf b}) Case 2, ({\bf c}) Case 3, ({\bf d}) Case 4, ({\bf e}) Case 5, ({\bf f}) Case 6, ({\bf g}) Case 7, ({\bf h}) Case 8, and ({\bf i}) Case 9.}
	\label{fig:shah-sens}
\end{figure}

The corresponding binary representation of the hash values are illustrated in Fig. \ref{fig:shah-sens}.

The result shows that the proposed hash function based on irregularly decimated chaotic map has high sensitivity to any changes to its security key space.
Even tiny changes in secret keys or in input messages will lead to significant differences of hash values.

\subsection{Statistic Analysis of Diffusion and Confusion}
From a historical point of view, Shannon, with the publication in 1949 of his paper, \textit{Communication Theory of Secrecy Systems} \cite{Shannon1949}, introduced the idea of two methods for frustrating a statistical analysis of encryption algorithms: confusion and diffusion. 

Confusion is intended to use transformations to hide the relationship between the plaintext and ciphertext, which means the relationship between the plaintext and the hash value is as complicated as possible.

Diffusion can propagate the change over the whole encrypted data, which means that the hash value is highly dependent on the plaintext. For a binary representation of the hash value, each bit can be only 0 or 1. 
Therefore, the ideal diffusion effect should be that any
tiny changes in the initial condition lead to a 50\% changing probability of each bit of hash value.

Six statistics used here are: 
minimum number of changed bits $B_{min}$,
maximum number of changed bits: $B_{max}$, 
mean changed bit number $\bar{B}$,
mean changed probability $P$,
standard deviation of the changed bit number $\Delta B$, and 
standard deviation $\Delta P$.

They are defined as follows:

Minimum number of changed bits: 
$B_{min}=min (\{ B_{i} \}_{i=1}^{N})$

Maximum number of changed bits: 
$B_{max}=max (\{ B_{i} \}_{i=1}^{N})$

Mean changed bit number: 
$\bar{B}=\frac{1}{N} \sum_{i=1}^{N}B_{i}$

Mean changed probability: 
$P=\frac{\bar{B}}{n} \times 100 \% $

Standard deviation of numbers of changed bits:\\
$\Delta B=\sqrt{\frac{1}{N-1} \sum_{i=1}^{N}(B_{i}-\bar{B})^{2}}$ 

Standard deviation: $\Delta P=\sqrt{\frac{1}{N-1} \sum_{i=1}^{N}(\frac{B_{i}}{n} -P)^{2}} \times 100 \%$, \\
where $N$ is the total number of tests and $B_{i}$ is the number of changed bits in the $i$-th test (Hamming distance).

Two types of statistical tests are performed: type A and type B. 
In the type A test, a random message, referred as the original message, of size $L=50n$ is generated and its corresponding $n-$bit hash value is computed.
Then, a new message is generated by choosing a single bit at random from the original message and modified to 0 if it is 1 or to 1 if it is 0.
The $n$-bit hash value of the new message is then compared with that of the original message and the Hamming distance between the two hash values is recorded as $B_{i}$.
This is then repeated $N$ times, where each time, a new original message is chosen and one of its bits is randomly chosen and modified to 0 if it is 1 or to 1 if it is 0.
Tables \ref{tab:typeA128}--\ref{tab:typeA1024} present results of these tests for $n=$ 128, 160, 256, 512, 1024.

\begin{table}[]
\caption{Statistical results for 128-bit hash values generated under tests of type A.}
\label{tab:typeA128}
\footnotesize
\centering
\begin{tabular}{lccccc}
\hline \hline
 	& $N$=256 & $N$=512  & $N$=1024  & $N$=2048 & $N$=10,000 \\
\hline
$B_{min}$       & 51 & 49 & 47 & 45 & 45 \\
$B_{max}$       & 80 & 89 & 89 & 89 & 89 \\
$\bar{B}$       & 64.5 & 63.998 & 64.11 & 63.99 & 64 \\
$P(\%)$         & 50.39 & 49.99 & 50.08 & 49.99 & 50.01 \\
$\Delta B$      & 5.43 & 5.66 & 5.31 & 5.51 & 5.6 \\
$\Delta P (\%)$ & 4.425 & 4.19 & 4.15 & 4.33 & 4.37 \\
\hline \hline
\end{tabular}
\end{table}

\begin{table}[]
\caption{Statistical results for 160-bit hash values generated under tests of type A.}
\label{tab:typeA160}
\footnotesize
\centering
\begin{tabular}{lccccc}
\hline \hline
 	& $N$=256 & $N$=512  & $N$=1024  & $N$=2048 & $N$=10,000 \\
\hline
$B_{min}$       & 61 & 59 & 56 & 56 & 56 \\
$B_{max}$       & 94 & 96 & 100 & 101 & 101 \\
$\bar{B}$       & 74.31 & 79.74 & 79.99 & 80 & 79.92 \\
$P(\%)$         & 49.57 & 49.84 & 49.99 & 50 & 49.95 \\
$\Delta B$      & 5.86 & 6.07 & 6.21 & 6.39 & 6.3 \\
$\Delta P (\%)$ & 3.66 & 3.79 & 3.88 & 3.99 & 3.94 \\
\hline \hline
\end{tabular}
\end{table}

\begin{table}[]
\caption{Statistical results for 256-bit hash values generated under tests of type A.}
\label{tab:typeA256}
\footnotesize
\centering
\begin{tabular}{lccccc}
\hline \hline
 	& $N$=256 & $N$=512  & $N$=1024  & $N$=2048 & $N$=10,000 \\
\hline
$B_{min}$       & 103 & 102 & 102 & 102 & 92 \\
$B_{max}$       & 148 & 152 & 156 & 161 & 162 \\
$\bar{B}$       & 128.01 & 127.95 & 127.95 & 127.94 & 128.04 \\
$P(\%)$         & 50 & 49.98 & 49.98 & 49.97 & 50.01 \\
$\Delta B$      & 8.04 & 8.1 & 7.99 & 7.94 & 7.99 \\
$\Delta P (\%)$ & 3.14 & 3.16 & 3.12 & 3.1 & 3.12 \\
\hline \hline
\end{tabular}
\end{table}

\begin{table}[]
\caption{Statistical results for 512-bit hash values generated under tests of type A.}
\label{tab:typeA512}
\footnotesize
\centering
\begin{tabular}{lccccc}
\hline \hline
 	& $N$=256 & $N$=512  & $N$=1024  & $N$=2048 & $N$=10,000 \\
\hline
$B_{min}$       & 214 & 214 & 214 & 214 & 212 \\
$B_{max}$       & 287 & 287 & 287 & 293 & 302 \\
$\bar{B}$       & 255.94 & 255.95 & 256.03 & 255.74 & 256.04 \\
$P(\%)$         & 49.99 & 49.99 & 50.01 & 49.95 & 50 \\
$\Delta B$      & 12.73 & 11.84 & 11.48 & 11.44 & 11.33 \\
$\Delta P (\%)$ & 2.48 & 2.31 & 2.24 & 2.23 & 2.21 \\
\hline \hline
\end{tabular}
\end{table}

\begin{table}[]
\caption{Statistical results for 1024-bit hash values generated under tests of type A.}
\label{tab:typeA1024}
\footnotesize
\centering
\begin{tabular}{lccccc}
\hline \hline
 	& $N$=256 & $N$=512  & $N$=1024  & $N$=2048 & $N$=10,000 \\
\hline
$B_{min}$       & 471 & 469 & 464 & 454 & 448 \\
$B_{max}$       & 561 & 561 & 561 & 563 & 577 \\
$\bar{B}$       & 513.7 & 512.73 & 512.25 & 511.79 & 511.97 \\
$P(\%)$         & 50.16 & 50.07 & 50.02 & 49.98 & 49.99 \\
$\Delta B$      & 15.42 & 15.37 & 15.47 & 15.61 & 15.85 \\
$\Delta P (\%)$ & 1.5 & 1.5 & 1.51 & 1.52 & 1.54 \\
\hline \hline
\end{tabular}
\end{table}

Comparing variables with most of these existing hash algorithms given in Table \ref{tab:typeA128compN10000}, the SHAH has small $\Delta B$ and $\Delta P$ values, respectively.

\begin{table}[]
\caption{Comparison of statistical results for 128-bit hash values and $N$=10,000, under tests of type A.}
\label{tab:typeA128compN10000}
\footnotesize
\centering
\begin{tabular}{lcccccc}
\hline \hline
 	& $B_{min}$ & $B_{max}$  & $\bar{B}$  & $P(\%)$ & $\Delta B$ & $\Delta P (\%)$ \\
\hline
SHAH 
& 45 & 89 & 64 & 50.01 & 5.6 & 4.37 \\
Ref. \cite{KansoGhebleh2013} 
& 45 & 84 & 63.94 & 49.95 & 5.64 & 4.41 \\
Ref. \cite{KansoGhebleh2015}
& 44 & 84 & 64 & 50 & 5.65 & 4.41 \\
Ref. \cite{KansoYahyaouiAlmulla2012} 
& 46 & 82 & 64.15 & 50.12 & 5.74 & 4.48 \\
Ref. \cite{LinYuLu2017} 
& 44 & 84 & 63.95 & 49.96 & 5.62 & 4.39 \\
Ref. \cite{WangWongXiao2011} 
& 42 & 83 & 63.986 & 49.988 & 5.616 & 4.388 \\
\hline \hline
\end{tabular}
\end{table}

In the type B test, the original message $M$ of size $ L=50n$ bits is  generated at random and its corresponding $n$-bit hash value is computed.
Then, a single bit of the original message is chosen, modified to 0 if it is 1 or to 1 if it is 0, and the hash value of the modified message is calculated.
The two hash values are compared, and the number of flipped bits is calculated and recorded as $B_{i}$.
The same original message is used for all $N$ iterations. Tables \ref{tab:typeB128}--\ref{tab:typeB1024} list the results obtained in tests of type B for $n=128, 160, 256, 512, 1024$, and various values of $N$.


\begin{table}[]
\caption{Statistical results for 128-bit hash values generated under tests of type B.}
\label{tab:typeB128}
\footnotesize
\centering
\begin{tabular}{lccccc}
\hline \hline
 	& $N$=256 & $N$=512  & $N$=1024  & $N$=2048 & $N$=50$\times$128  \\
\hline
$B_{min}$       & 53 & 49 & 48 & 43 & 43 \\
$B_{max}$       & 78 & 83 & 83 & 83 & 84 \\
$\bar{B}$       & 63.57 & 64.01 & 63.97 & 64.11 & 63.9 \\
$P(\%)$         & 49.66 & 50 & 49.97 & 50.08 & 49.92 \\
$\Delta B$      & 5.19 & 5.61 & 5.6 & 5.59 & 5.71 \\
$\Delta P (\%)$ & 4.06 & 4.38 & 4.37 & 4.37 & 4.46 \\
\hline \hline
\end{tabular}
\end{table}

\begin{table}[]
\caption{Statistical results for 160-bit hash values generated under tests of type B.}
\label{tab:typeB160}
\footnotesize
\centering
\begin{tabular}{lccccc}
\hline \hline
 	& $N$=256 & $N$=512  & $N$=1024  & $N$=2048 & $N$=50$\times$160 \\
\hline
$B_{min}$       & 66 & 62 & 60 & 57 & 57 \\
$B_{max}$       & 94 & 96 & 101 & 101 & 108 \\
$\bar{B}$       & 80.06 & 80.64 & 79.94 & 80.52 & 80.29 \\
$P(\%)$         & 50.03 & 50.4 & 49.96 & 50.32 & 50.18 \\
$\Delta B$      & 6.17 & 6.16 & 6.35 & 6.66 & 6.53 \\
$\Delta P (\%)$ & 3.85 & 3.85 & 3.96 & 4.16 & 4.08 \\
\hline \hline
\end{tabular}
\end{table}

\begin{table}[]
\caption{Statistical results for 256-bit hash values generated under tests of type B.}
\label{tab:typeB256}
\footnotesize
\centering
\begin{tabular}{lccccc}
\hline \hline
 	& $N$=256 & $N$=512  & $N$=1024  & $N$=2048 & $N$=50$\times$256 \\
\hline
$B_{min}$       & 108 & 107 & 106 & 104 & 100 \\
$B_{max}$       & 143 & 148 & 151 & 153 & 156 \\
$\bar{B}$       & 128.23 & 129.23 & 128.6 & 128.98 & 128.26 \\
$P(\%)$         & 50.09 & 50.47 & 50.23 & 50.38 & 50.1 \\
$\Delta B$      & 8.44 & 8.12 & 8.05 & 8.08 & 8.06 \\
$\Delta P (\%)$ & 3.29 & 3.17 & 3.14 & 3.15 & 3.14 \\
\hline \hline
\end{tabular}
\end{table}

\begin{table}[]
\caption{Statistical results for 512-bit hash values generated under tests of type B.}
\label{tab:typeB512}
\footnotesize
\centering
\begin{tabular}{lccccc}
\hline \hline
 	& $N$=256 & $N$=512  & $N$=1024  & $N$=2048 & $N$=50$\times$512 \\
\hline
$B_{min}$       & 229 & 229 & 229 & 217 & 216 \\
$B_{max}$       & 278 & 292 & 292 & 300 & 300 \\
$\bar{B}$       & 255.28 & 255.82 & 256.45 & 256.62 & 256.11 \\
$P(\%)$         & 49.86 & 49.96 & 50.08 & 50.12 & 50.02 \\
$\Delta B$      & 11.95 & 11.42 & 11.32 & 11.29 & 11.31 \\
$\Delta P (\%)$ & 2.33 & 2.23 & 2.21 & 2.2 & 2.2 \\
\hline \hline
\end{tabular}
\end{table}

\begin{table}[]
\caption{Statistical results for 1024-bit hash values generated under tests of type B.}
\label{tab:typeB1024}
\footnotesize
\centering
\begin{tabular}{lccccc}
\hline \hline
 	& $N$=256 & $N$=512  & $N$=1024  & $N$=2048 & $N$=50$\times$1024 \\
\hline
$B_{min}$       & 473 & 472 & 464 & 458 & 441 \\
$B_{max}$       & 551 & 556 & 556 & 557 & 568 \\
$\bar{B}$       & 510.56 & 511.32 & 512.41 & 512.32 & 511.61 \\
$P(\%)$         & 49.85 & 49.93 & 50.04 & 50.03 & 49.96 \\
$\Delta B$      & 13.7 & 14.82 & 15.16 & 15.85 & 15.95 \\
$\Delta P (\%)$ & 1.33 & 1.44 & 1.48 & 1.54 & 1.55 \\
\hline \hline
\end{tabular}
\end{table}

Comparing the results with few chaos based hash algorithms given in Table \ref{tab:typeB128compN2048}, the SHAH has small $\Delta B$ and $\Delta P$ values, accordingly.

\begin{table}[]
\caption{Comparison of statistical results for 128-bit hash values and $N$=2048, under tests of type B.}
\label{tab:typeB128compN2048}
\footnotesize
\centering
\begin{tabular}{lcccccc}
\hline \hline
 	& $B_{min}$ & $B_{max}$  & $\bar{B}$  & $P(\%)$ & $\Delta B$ & $\Delta P (\%)$ \\
\hline
SHAH 
& 43 & 83 & 64.11 & 50.08 & 5.59 & 4.37 \\
Ref. \cite{KansoGhebleh2013}
& 47 & 81 & 63.95 & 49.96 & 5.62 & 4.39 \\
Ref. \cite{KansoGhebleh2015}
& 48 & 83 & 64.22 & 50.17 & 5.65 & 4.42 \\
Ref. \cite{KansoYahyaouiAlmulla2012} 
& 47 & 84 & 63.94 & 49.95 & 5.69 & 4.44 \\
\hline \hline
\end{tabular}
\end{table}

In Tables \ref{tab:typeA128}--\ref{tab:typeB1024} we can observe that both types of tests, the mean changed bit number $\bar{B}$ and the mean probability $P$ are very close to the ideal values $n/2$ and $50\%$. These results indicate that the suggested hashing scheme has very robust capability for confusion and diffusion. Thus, the SHAH function is trustworthy against this type of attacks.

\subsection{Collision Analysis}
In this section we will analyse the novel hash function SHAH based on the collision tests proposed in \cite{KansoGhebleh2013}.
In general, a common characteristic of a hash scheme is to have a  collision resistance capability, the following two types of tests are performed, type A and type B.
In tests of type A, an input message of size $L=50n$ is generated and its corresponding $n-$bit hash value is computed and stored in ASCII format.
Then, a new message is generated by choosing a single bit at random from the input message and modified to 0 if it is 1 or to 1 if it is 0.
The $n$-bit hash value of the new message is calculated and stored in ASCII format.
The two hash values are compared, and the number of ASCII symbols with the same value at the same location is counted.
Moreover, the absolute difference $D$ between the two hash values is computed by the following formula
$D = \sum_{i=1}^{n/8}|dec(e_i)-dec(e_i^{'})|$, 
where $e_i$ and $e_{i}^{'}$ be the $i$-th entry of the input and new hash value, respectively, and function $dec()$ converts the entries to their equivalent decimal values.
The test of type A is repeated $N=10,000$ times, and experimental minimum, maximum, and mean of $D$ are presented in Table \ref{tab:typeACollision} for different hash values of size $n=$ 128, 160, 256, and 512.

\begin{table}[]
\caption{Absolute difference $D$ for hash values generated under tests of type A, where $N=10,000$.}
\label{tab:typeACollision}
\footnotesize
\centering
\begin{tabular}{lccc}
\hline \hline
 $n$	& Maximum & Minimum  & Mean \\
\hline
128       & 2386 & 537 & 1367 \\
160       & 2821 & 717 & 1706 \\
256       & 4049 & 1395 & 2731 \\
512       & 7517 & 3922 & 5459\\
\hline \hline
\end{tabular}
\end{table}

Table \ref{tab:typeACollisionComp128} outlines the absolute differences of 128-hash values generated under tests of type A, where $N=10,000$, of some existing hash functions which are based on chaotic maps. The results show that the SHAH has comparable values. 

\begin{table}[]
\caption{Comparison of the absolute difference for 128-hash values generated under tests of type A, where $N=10,000$.}
\label{tab:typeACollisionComp128}
\footnotesize
\centering
\begin{tabular}{lccc}
\hline \hline
 $n$	& Maximum & Minimum  & Mean \\
\hline
SHAH
& 2386 & 537 & 1367 \\
Ref. \cite{KansoGhebleh2013}
& 2391 & 656 & 1364 \\
Ref. \cite{KansoYahyaouiAlmulla2012}
& 2320 & 737 & 1494 \\
Ref. \cite{{RenWangXieYang2009}}
& 2455 & 599 & 1439 \\
Ref. \cite{WangWongXiao2011}
& 2064 & 655 & 1367\\
\hline \hline
\end{tabular}
\end{table}

\begin{figure}[ht]
\begin{center}
	\includegraphics[height=.25\textheight]{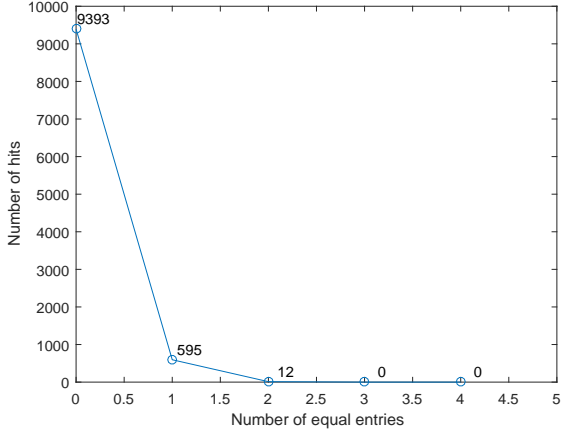}
\end{center}
	\caption{Distribution of the number of locations where the ASCII symbols are equal in the 128-bit hash values generated under tests of type A, where
$N=10,000$}
	\label{fig:CollTestA128}
\end{figure}

The number of hits where the ASCII symbols are equal, where $N=10,000$ and the hash values are generated under tests of type A, is listed in Table \ref{tab:typeACollisionEqual10} and distribution of the 128-hash values, is presented in Figure \ref{fig:CollTestA128}.

\begin{table}[]
\caption{Count of hits in Collision test for $N=10,000$, generated under tests of type A.}
\label{tab:typeACollisionEqual10}
\footnotesize
\centering
\begin{tabular}{lcccccc}
\hline \hline
 $n$	& 0 & 1  & 2 & 3 & 4 & 5\\
\hline
128       & 9393 & 595 & 12 & 0 & 0 & 0 \\
160       & 9236 & 734 & 30 & 0 & 0 & 0 \\
256       & 8795 & 1138 & 64 & 3 & 0 & 0 \\
512       & 7789 & 1944 & 242 & 23 & 0 & 0 \\
\hline \hline
\end{tabular}
\end{table}

In the type B tests, an input message $M$ of a fixed size $ L=50n$ bits is  created at random and its corresponding $n$-bit input hash value is computed.
Then, a single bit of the input message is chosen, modified to 0 if it is 1 or to 1 if it is 0, and the hash value of the modified message is calculated.
Table \ref{tab:typeBCollision} presents minimum, maximum, and mean values of $D$ for different hash values of size $n = 128, 160, 256,$ and $512$ generated under tests of type B.
The same input message is used for all $N$ iterations. Comparison with other algorithm is presented in Table \ref{tab:typeBCollision128Comp}.

\begin{table}[]
\caption{Absolute difference $D$ for hash values generated under tests of type B, where $N=50n$.}
\label{tab:typeBCollision}
\footnotesize
\centering
\begin{tabular}{lcccc}
\hline \hline
 $n$	& Maximum & Minimum  & Mean & $N$\\
\hline
128       & 2035 & 636 & 1248 & 6400\\
160       & 2830 & 969 & 1908 & 8000\\
256       & 4124 & 1483 & 2817 & 12800\\
512       & 7469 & 3726 & 5709 & 25600\\
\hline \hline
\end{tabular}
\end{table}

\begin{table}[]
\caption{Comparison of absolute difference $D$ for 128-hash values generated under tests of type B, where $N=6400$.}
\label{tab:typeBCollision128Comp}
\footnotesize
\centering
\begin{tabular}{lccc}
\hline \hline
 $n$	& Maximum & Minimum  & Mean \\
\hline
SHAH
& 2035 & 636 & 1248 \\
Ref. \cite{KansoGhebleh2013}
& 2421 & 735 & 1576 \\
Ref. \cite{KansoGhebleh2015}
& 2294 & 661 & 1360 \\
\hline \hline
\end{tabular}
\end{table}

\begin{figure}[ht]
\begin{center}
	\includegraphics[height=.25\textheight]{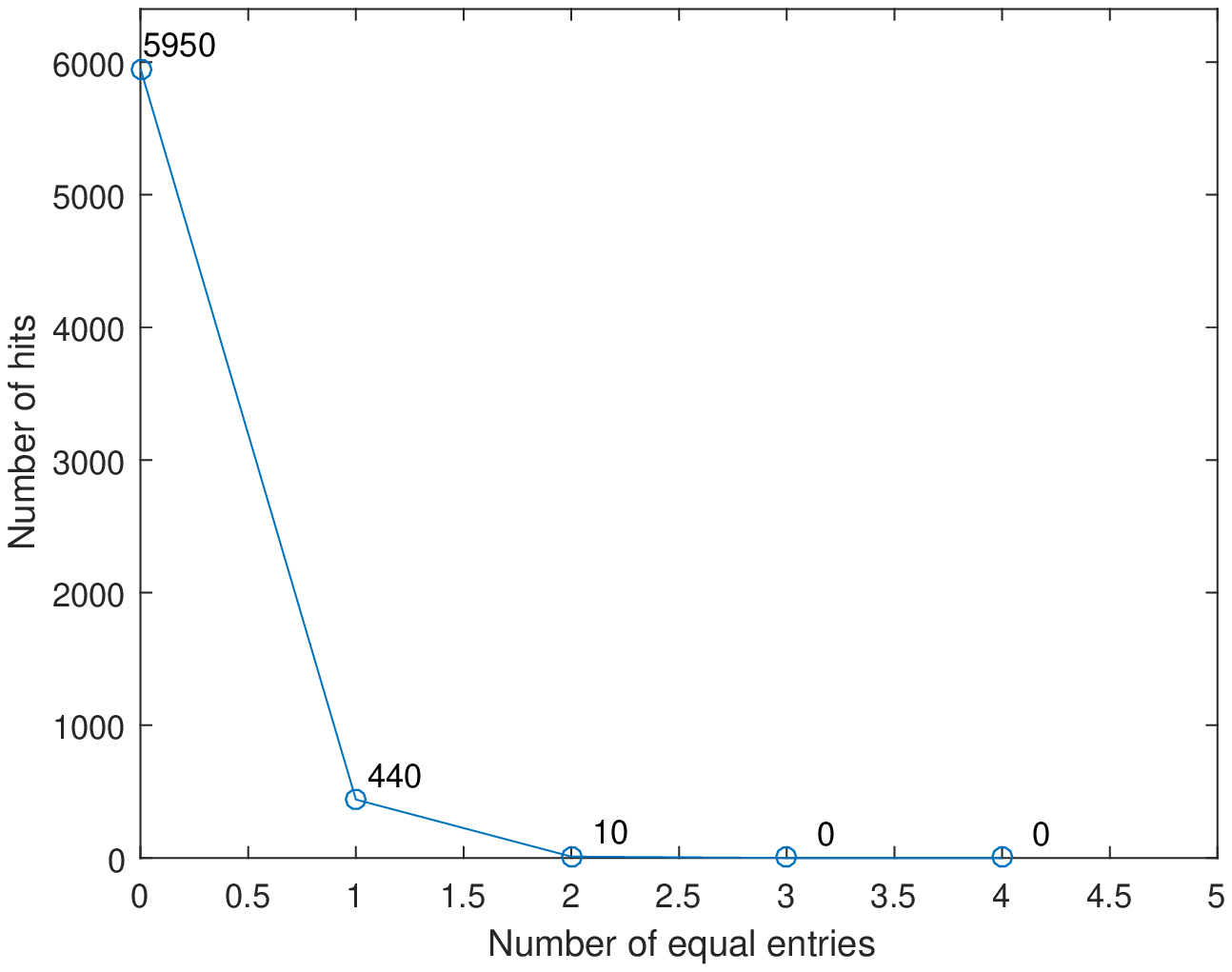}
\end{center}
	\caption{Distribution of the number of locations where the ASCII symbols are equal in the 128-bit hash values generated under tests of type B, where
$N=6400$}
	\label{fig:CollTestB6400}
\end{figure}

Distribution of the number of locations where the ASCII symbols are equal in the 128-bit hash values generated under tests of type B, where $N=50 \times 128 = 6400$ are presented in Figure \ref{fig:CollTestB6400}.

In addition to the above experiments, the tests of type B are repeated for very short input strings consisting of a single $n-bit$
block, Table \ref{tab:typeBCollisionShort}.

\begin{table}[]
\caption{Absolute difference $D$ for hash values generated under tests of type B, where $N=n$.}
\label{tab:typeBCollisionShort}
\footnotesize
\centering
\begin{tabular}{lcccc}
\hline \hline
 $n$	& Maximum & Minimum  & Mean & $N$\\
\hline
128       & 1812 & 844  & 1263 & 128\\
160       & 2480 & 1253 & 1840 & 160\\
256       & 3894 & 2065 & 2843 & 256\\
512       & 7469 & 4400 & 5695 & 512\\
\hline \hline
\end{tabular}
\end{table}

\begin{figure}[ht]
\begin{center}
	\includegraphics[height=.25\textheight]{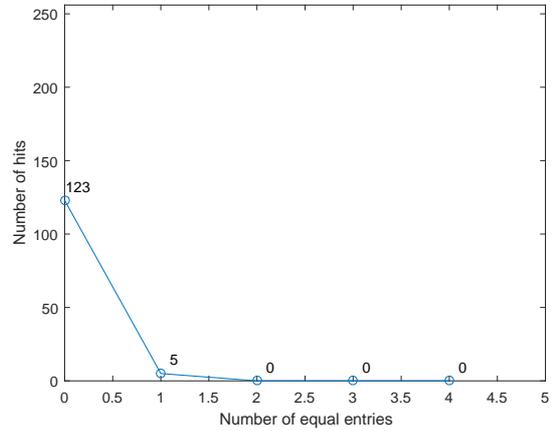}
\end{center}
	\caption{Distribution of the number of locations where the ASCII symbols are equal in the 128-bit hash values generated under tests of type B, where
$N=128$ and $L=n$}
	\label{fig:CollTestB128}
\end{figure}

Distribution of the number of locations where the ASCII symbols are equal in the 128-bit hash values generated under tests of type B, where $N=128$ and $L=n$ are presented in Figure \ref{fig:CollTestB128}.

From the obtained results it is clear that the novel hash function SHAH has a strong collision resistance capacity. Compared with similar hash functions, the proposed one has  a mean per character values close to the ideal of 85.3333 \cite{DengLiXiao2010} and low collision values.


\section{Conclusions}
In this paper, we propose a novel hash function based on shrinking chaotic map. The hash function called SHAH is based on two Tinkerbell maps filtered with the decimation rule.
Exact study has been provided on the novel scheme using distribution analysis, sensitivity analysis, static analysis of diffusion and confusion, and collision analysis. The experimental data show excellent performance of the SHAH algorithm.


%

\section*{Acknowledgment}

The authors are grateful to the anonymous referees for valuable and helpful comments.

This work is supported by the Scientic research fund of Konstantin Preslavski University of Shumen under the grant No.  RD-08-121\slash 06.02.2018 and by European Regional Development Fund and the Operational Program "Science and Education for Smart Growth" under contract UNITe No. BG05M2OP001-1.001-0004-C01 (2018-2023).

\ifCLASSOPTIONcaptionsoff
  \newpage
\fi

\end{document}